%% file: aaskaii_maser_astroph.tex
\documentclass[a4paper,11pt]{article}
\usepackage{aaskaiid}
\usepackage{orcidlink}
\usepackage{soul}
\newcommand{\virg}[1]{``#1''}   
\newcommand{\unavirg}[1]{`#1'}
\title{Boosting Water Maser Studies in AGN with the SKA}
\ShortTitle{Water Masers in AGN with the SKA}

\author[1]{Andrea Tarchi\orcidlink{0000-0001-8540-3500}}
\ShortName{A. Tarchi et al.} 
\author[1]{Paola Castangia\orcidlink{0000-0002-2318-8531}}
\author[1]{Elisabetta Ladu\orcidlink{0009-0005-5980-0795}}

\affiliation[1]{INAF-Osservatorio Astronomico di Cagliari, via della Scienza 5, 09047, Selargius (CA), Italy}
\emailAdd{andrea.tarchi@inaf.it}
\emailAdd{paola.castangia@inaf.it}
\emailAdd{elisabetta.ladu@inaf.it}

\abstract{Extragalactic water maser sources are unique tools to derive fundamental physical quantities of
the host galaxies. In nearby and distant active galactic nuclei (AGN), water masers are used to determine the
geometry of accretion disks around super-massive black holes, precise black hole masses, and
standard-candles-independent distances to the host galaxy. In addition, they allow detailed studies
of the interaction between nuclear jets/outflows and the interstellar medium, providing clues on AGN feedback mechanisms.
So far, however, extragalactic maser searches have yielded detection rates of few percent, and only
relatively few maser sources have been found, mostly in the nearby Universe.
Because of its unprecedented sensitivity, the SKA will allow to significantly increase the number of known water
maser sources especially in the more distant Universe. This will lead to the chance of performing statistically-relevant studies of the
maser phenomenon (and its occurrence), derive extragalactic maser luminosity functions 
and, ultimately, to perform the aforementioned studies for larger samples and up to cosmological distances.
In this Chapter, we will provide a quantitative analysis of the expected number of new
extragalactic water maser sources already at the reach of the SKA-Mid telescope (in AA4 configuration) through targeted and blinds surveys. 
In addition, we will discuss the main requirements for the upcoming SKA design, in terms of baselines and frequency coverage, that may maximize the exploitation of such wealth of new targets, allowing a true step forward in AGN-related maser science.}

\begin{document}
\maketitle
\include{journal-names}

\section{Introduction}

A fundamental element of the Unified Model of AGN is the dusty toroidal structure, known as the \virg{torus} (e.g.,  \citealt{Antonucci1993}; \citealt{Urry95}), that surrounds the super massive black holes (SMBH). Such a structure is thought to block the direct emission from the accretion disk, scattering and reemitting it in the infrared (IR). In the last 10-15 years, IR and X-ray studies of AGN provided significant information of the obscuring material in the proximity of SMBH. As a consequence, the paradigmatic homogeneous dusty torus, seen as an isolated component, has been replaced by a discrete, \virg{clumpy} structure, physically and dynamically connected with the host galaxy through gas inflows/outflows (e.g., \citealt{Almeida2017}). However, despite the great advances done, so far, the accurate geometry and dynamical origin of the absorbing material have not yet been fully understood. In addition, further importance for a better knowledge of the AGN phenomenon is represented by the yet debated accretion/ejection mechanisms taking place in the nuclear regions of radio-loud and, even more, radio-quiet AGN. The aforementioned studies are, however, made complicated by the very small linear dimensions and by the fact that the internal regions are often obscured at optical and UV wavelengths. IR, X-rays , and radio observations can, nevertheless, access such obscured regions.

In this Chapter, we will focus on the contribution of the 22-GHz water maser radio spectral lines in AGN to a better understanding of the aforementioned issues\footnote{A description of the potential of the SKA-Mid telescope for studies of stellar and interstellar masers in nearby galaxies is addressed in \cite{Rygl01.2026.SKA}, this volume}, and the chances offered by the use of SKA-Mid in its AA4 configuration to significantly increase the number of known sources up to cosmological distances. The range in frequencies covered by the 5a and 5b bands are suitable to detect the 22-GHz water maser line in galaxies with redshift, $z$, between 0.45 and 3.8. However, the possibility to extend the actual project design to higher frequency is also discussed, together with a remark on the importance of the availability of (very) long baselines for detailed follow-ups of individual sources. 

Indeed, the radio emission from luminous H$_2$O masers, the so-called \virg{megamasers} constitutes the only way to directly map the molecular gas at pc/sub-pc distance from SMBHs (for additional details on masers in AGN, please see the reviews by, e.g., \cite{Lo2005}, \cite{Henkel2005}, \cite{Tarchi2012}, \cite{Pesce2018}. 

Some maser sources, labeled as \virg{disk-masers}, are associated with nuclear accretion disks or sometimes with the innermost boundary of a dusty torus in the region of the interaction with the outflows (e.g., \citealt{Bannikova2023}).
Very Long baseline Interferometry (VLBI) and single-dish monitoring studies of these masers can be used to determine the geometry of accretion disks and to estimate the enclosed dynamical masses and provide a calibration of the cosmic distance scale (\citealt{Miyoshi1995}; \citealt{Herrnstein1997}; \citealt{Kuo2011}; \citealt{Gao2017}). H$_2$O masers currently provide the most precise method to determine black hole masses in external galaxies, especially in the case of type 2 AGN, where other methods, such as those based on optical broad lines emission, cannot be used. Measuring black hole masses in AGN, allows one to estimate Eddington luminosities and accretion efficiencies. In addition, these studies are extremely relevant to probe the low mass regime of the M$_{BH}$ -- $\sigma^{*}$ relation, so far, almost uniquely derived for elliptical galaxies with larger BH masses (e.g., \citealt{Greene2010}).

Water maser emission can also be associated with radio jets, produced by either the interaction between the radio jet and an encroaching molecular cloud or by the amplification of the radio continuum from the jet from excited water molecules in a foreground cloud. Detailed studies of these masers allow us to pinpoint regions of strong interaction of the jets with the interstellar medium of the host galaxy (e.g., \citealt{Castangia2019}) and derive relevant physical quantities of the jet material, like its velocity and density (e.g., \citealt{Peck2003}).

A third class of AGN-associated water masers is named \virg{outflow-masers}. They trace the velocity and geometry of nuclear winds at $\leq$ 1 pc from the nucleus, as in the case of Circinus (\citealt{Greenhill2003}) and NGC 3079 (\citealt{Kondratko2005}), offering a promising means to probe the structure and motion of the clouds in the toroidal obscuring region predicted by clumpy torus models (e.g., \citealt{Nenkova2008}), and to help studies of AGN tori in general.


Each megamaser source is therefore a goldmine of information on the (sub)-parsec-scale environment around AGN, making the discovery of new sources and their interferometric follow-ups crucial for their study. So far, however, among the $\geq$ 6000 galaxy nuclei surveyed for 22-GHz water maser emission, only 180 sources are detected, with $\sim$ 30\% of them possibly originating in disc (\citealt{Kuo2020}), and the vast majority being hosted in radio-quiet AGN, mainly Seyfert\,2s and LINERs, in the local Universe (z $<$ 0.05). Indeed, the combination of the high sensitivity required in surveys and the large number of objects to be searched for maser emission has played a role against a more conspicuous number of detections.
The NRAO Key Science Megamaser Cosmology Project (MCP; \citealt{Reid2009}), for example, observed about 3000 obscured AGN (Seyfert 2), mainly selected from large optical surveys, such as SDSS, 6dF and 2MRS, detecting water maser emission in only 3\% of the targets (\citealt{Braatz2018}). Higher detection rates have been obtained by selecting galaxies on the basis of their IR and X-ray properties. In particular, a sample of galaxies with IRAS point source flux density $>$50\,Jy, yielded a maser detection rate of 23\% (\citealt{Henkel2005}; \citealt{Surcis2009}). More recently, \cite{Panessa2020} obtained a detection rate of $\sim$20\% in a complete sample of AGN selected in the 20--40\,keV energy range. Through a combination of mid-IR (\textit{IRAS}) and X-ray (\textit{XMM-Newton}) data, instead, \cite{severgnini2012} selected a well defined sample of 36 heavily absorbed AGN ($N_{\rm H} \geq$10$^{23}$ cm$^{-2}$) that have been searched for 22\,GHz water maser emission. The maser detection rate of this sample is a remarkable 50\%, one of the highest ever obtained in extragalactic maser searches (\citealt{Castangia2019}).
No water maser source has been reported, so far, at redshifts larger than 0.1, with the exception of two cases, the type 2 QSO J0804+3607 (at $z$=0.66; \citealt{Barvainis2005}) and the grativationally-lensed type 1 quasar J0414+0534 (at $z$=2.64, \citealt{Impellizzeri2008}) .

\section{Water masers at cosmological redshifts with the SKA-Mid in AA4 Configuration}

The discovery of a water maser in the gravitational lens MG\,J0414$+$0534 at $z$=2.64 (\cite{Impellizzeri2008}), opened the door for the study of H$_2$O masers at high redshifts. Indeed, using the magnification provided by the foreground gravitational lens to increase the observed flux density and the angular extent of any water maser in the background, it is in principle possible to map parsec-scale accretion disks in distant AGN, also with current instrumentation. Furthermore, this discovery suggests that the space density of luminous water masers was larger at high redshift than in the local Universe (\citealt{Impellizzeri2008}). 

Following the detection of the maser in the gravitational lens MG\,J0414$+$0534 at $z$=2.64 (\citealt{Impellizzeri2008}), \cite{Mckean2011} (hereafter M11) performed a search for water maser emission in a small sample of (five) dusty, gravitationally lensed quasars and star-forming galaxies at redshifts between 2.3 and 2.9. Despite no new confident detection (above the 3$\sigma$ level) was found, this study allowed to update the water maser luminosity function (LF) derived by \cite{Henkel2005} and \cite{Bennert2009} at high redshift. In particular, M11 demonstrates that there must be some evolution in the LF at moderate redshifts, thus motivating high-sensitivity blind surveys for the water maser transition at high $z$. Some results from the {\it Herschel} space observatory also suggest that high-z ultra-luminous infrared galaxies tend to be very strong emitters in water vapor (\citealt{Omont2013}), reinforcing the motivation to search for H$_2$O masers at cosmological distances. Applying the reasoning of M11, with all the reported assumptions and caveats, by parametrizing the evolution of the LF with redshift, i.e. $(1+z)^m$, with the parameter $m$ equal to 0, 4 or 8, to account for a no, moderate, and strong evolution, respectively, we can predict the number of water maser galaxies found in pointed and/or blind surveys for a certain survey detection limit.
Indeed, with the advent of the next generation of radio telescopes, such as the SKA and the ngVLA, it will become straightforward to perform either blind surveys of large areas in the sky and/or deep targeted surveys of large samples of selected galaxies up to cosmological redshifts.

{\it Targeted searches:} In this framework, the SKA-Mid will be able to obtain significant results by performing targeted searches for water maser lines of extremely large samples of galaxies. The same spectral-line sensitivity of a 100-m class single-dish (i.e., the antennas that are typically involved in water maser searches) will be reached by SKA-Mid in a factor 10 less time, thus allowing to search (with comparable sensitivity) much larger samples and/or providing detections of a still significant number of maser sources also at larger distances ($z$ $>$ 0.5). These latter sources will likely be detected among the brighter members of the megamaser class (a 30-minutes observation would detect a 250 solar luminosity source at redshift of 0.5). Moreover, any such detection will yield the potential to detect and explore the maser phenomenon, and the AGN activity at which it is associated, in classes of galaxies where, so far, no maser have been found (e.g., radio loud objects, typically located at larger distances). Deeper targeted survey will then allow to push further the detection limit also to cosmological redshifts. Indeed, using 10-h long measurements, will potentially yield detections of maser sources with 1000 solar luminosities out to redshift 2.

{\it Blind searches:} as shown by M11, when approaching cosmological distances $z$ $\ge$ 1, blind surveys of water maser sources with SKA-Mid will also become valuable, particularly if, as expected, an evolution of the water maser LF is present. Under this latter assumption, as shown in  Fig.~1\footnote{The plot was produced by adapting the computational steps reported by \cite{Darling2002} for OH megamasers (their Eq.~18) for the water maser case.}, also blind searches with the SKA-Mid array would yield, at $z$ $\ge$ 0.45 (the lowest redshift observable in the 22-GHz maser line at the frequency of 15.4 GHz) , a number of new maser detections between $\sim$ 170 and 14000, for a LF evolution exponent of 4 and 8, respectively, when areas of the order of a square degree (about 65 pointings, considering the field of view of the array at 10 GHz) are searched for. This can be attained with a sensitivity (3$\sigma$)\footnote{The reported values are computed using the SKA-Mid sensitivity calculator, available at: \unavirg{https://sensitivity-calculator.skao.int/}, for a compromise frequency of 10 GHz.} of $\sim$ 0.06 mJy for a 10-km/s channel (reachable in about 10 hours on-source per pointing, and hence, a total of $\sim$ 1000 hours for each of the three configurations necessary to cover the 5a and 5b bands). This is surely feasible thanks to the enhanced sensitivity and the large field of view of the SKA array, and together with the large instantaneous bandwidths (3.9 and 2$\times$ 2.5 GHz for band 5a and 5b, respectively) available, that allow to cover at once relevant redshift ranges with suitable spectral resolution. At relatively low redshifts, blind surveys are less effective than targeted ones. However, they become increasingly relevant if and when large portions of the sky are surveyed with sufficient sensitivity. Since this may imply to invest a considerable amount of observing time, a good degree of commensality and complementarity with other projects would be ideal. In Fig. 2, we show a plot with the number of masers as a function of redshift ($z$ $\ge$ 0.45) for a quarter-sky search, with a total integration time of $\sim$ 1000 hours for each configuration (with integration time and 3-$\sigma$ sensitivity, per pointing, of 5 sec and 6 mJy/chan, respectively). Assuming an evolution of the LF with redshift (see above), in the aforementioned survey, at redshift between 0.45 and 1.2, we would expect to detect between $\sim$ 80 and 800 new maser sources, for a LF evolution exponent of 4 and 8, respectively.\\

To recapitulate: the use of the SKA-Mid telescope in the AA4 configuration will offer the unique chance to unveil hundreds to thousands of new water maser sources in a, so far, almost maser-undetected range of redshift (between 0.5 and 2), and, possibly, larger variety of objects.
While the use of SKA-Mid in its AA* configuration will already show promising avenues for water maser surveys, it has to be pointed out that the almost factor of two loss in sensitivity w.r.t. the AA4 configuration, will turn into a corresponding significant decrease by half in efficiency for targeted searches, and of a factor two and six in the number of maser detections in a square degree or quarter-sky blind searches, respectively. Independently from which surveying strategy (targeted or blind) will be adopted, it is evident that the inclusion of the SKA-Mid array, particularly in its AA4 configuration, in the search for water masers will be a game changer by definition, since it will permit the detection of many more sources in a part of the sky -- the southern one -- so far, largely unexplored or observed with shallow sensitivity. Indeed, the radioastronomical facilities that have access to the sky with declination below $-$40 degrees (e.g., the Murriyang 64-m radio telescope and the Australia Telescope Compact Array) are presently not sensitive enough to perform searches for extragalactic water masers, as the one described above, in a reasonable amount of time.
Noticeably, relevant complementary information on AGN detected in the 22-GHz water maser line can be derived from studies of water maser transitions at mm/sub-mm wavelengths recently made possible by the remarkable capabilities of ALMA and NOEMA (e.g., \citealt{Tarchi2024}, and references therein).\\

\section{Technical requirements and possible enhancements beyond the design baseline}

In the previous section, the capability for the SKA-Mid to detect new water maser sources has been quantitatively discussed. Indeed, a number of relevant details can be inferred from the spatially-unresolved spectra of the water maser emission alone. The spectral line profiles can in fact provide indications if the maser originates from an accretion disk, where three groups of narrow features centered at the systemic velocity of the host galaxy are observed, or in radio jets/outflows, where a single broad line displaced from the systemic velocity is often displayed (e.g., \citealt{Tarchi2012}). However, in order to maximize the return of any new maser detection and, in particular, to determine precise black-hole mass and distance estimates of the host galaxies in disk masers, high spatial resolution imaging capabilities are required to map the distribution of the water maser spots. 
While the prototypical disk-maser case is that in NGC\,4258 whose accretion disk has a subparsec scale size, more and more disk-maser sources have been and are under investigation, especially in the framework of the Megamaser Cosmology Project (see, e.g., \citealt{Braatz2010}). In some cases, disk sizes of the order of 1 pc have been also observed and, possibly, even larger accretion disks are expected in radiogalaxies were more massive black-holes are present (\citealt{Tarchi2007}). 
Hence, as shown in Fig. 3, a sensitive array with baselines of thousands of kms would open up the possibility to lead detailed studies of accretion disks and jets/outflows in AGN also in distant galaxies (up to $z \approx 1$ or more, if also space-VLBI baselines will become available in the future), for which mJy/sub-mJy sensitivities and sub-parsec/parsec linear resolutions are necessary. Therefore, a synergy between the SKA-Mid and the Global VLBI framework is then absolutely desirable.
Clearly, in order to resolve a 1-pc structure at $z$=1, at 11 GHz (the frequency at which the water maser line is Doppler-shifted), baselines  $>$10$^{5}$ km are required. Consequently, only maser sources (much) closer than $z$=1, masing structures with larger extent, or gravitationally-lensed maser galaxies (for such a case, see \citealt{Impellizzeri2008} and \citealt{Castangia2011}), will provide substantial advantage in obtaining relevant information of the AGN components through follow-up interferometric studies with earth-based antennas only. For these studies, a network of radiotelescopes is then recommended that combines the SKA-Mid array (possibly used in phased mode) with a number of antennas (or other suitable arrays, e.g., the ATCA, KVN, or VERA) at relatively-large distances to provide the necessary baseline lengths (few thousands of km). Instead, as shown before, in the case of unlensed water masers at cosmological distances, only space VLBI (presently represented by RadioAstron) may grant the necessary resolution for detailed studies of the emission.

Furthermore, the highest frequency presently planned for SKA-Mid is 15.4 GHz, thus preventing the exploitation of its enhanced capabilities also for searches and detailed studies of 22-GHz water maser sources in nearby ($z$$<$0.45) targets. This \virg{gap} will, however, be filled by taking profit of the future ngVLA project (at least for source declinations above $-$40 degrees) and/or by the possible extension of the SKA-mid observing frequency beyond band 5b. Overall, the joint effort of these instruments will allow us to significantly increase the number of nearby and distant water maser sources with the particularly appealing possibility, among others, to study, for the first time, these objects in radio loud galaxies, and to test the possible evolution of the maser LF with redshift.


\begin{figure}[htbp]
    \centering
    \includegraphics[trim=80 350 90 90, clip, width=1.0\columnwidth]{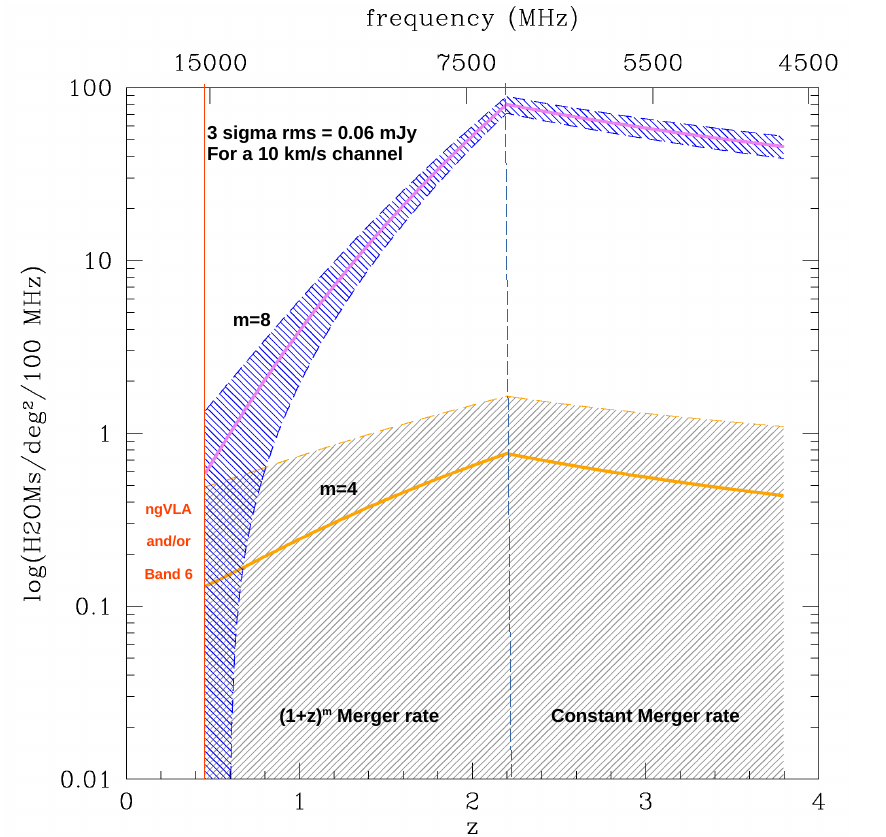}
   \caption{Sky density of detectable water masers with the SKA-Mid telescope (AA4) blind-searching a portion of sky of 1 square degree at a sensitivity of 0.06 mJy. Shaded areas indicate the 1$\sigma$ confidence level. The evolution of the LF with redshift is parametrized as $(1+z)^m$, where $m$ is 4 and 8 for moderate and very high evolution, respectively. The dashed line marks the redshift ($z$=2.2), after which there is assumed to be constant density evolution. The red solid line defines the actual lower limit in redshift ($z$ $\sim$ 0.45) for 22-GHz water maser studies possible with the SKA-Mid telescope due to the highest frequency threshold of band 5b (15.4 GHz).}
   \label{fig:1deg_figure}
\end{figure}

\begin{figure}[htbp]
    \centering
	\includegraphics[trim=80 350 90 90, clip, width=1.0\columnwidth]{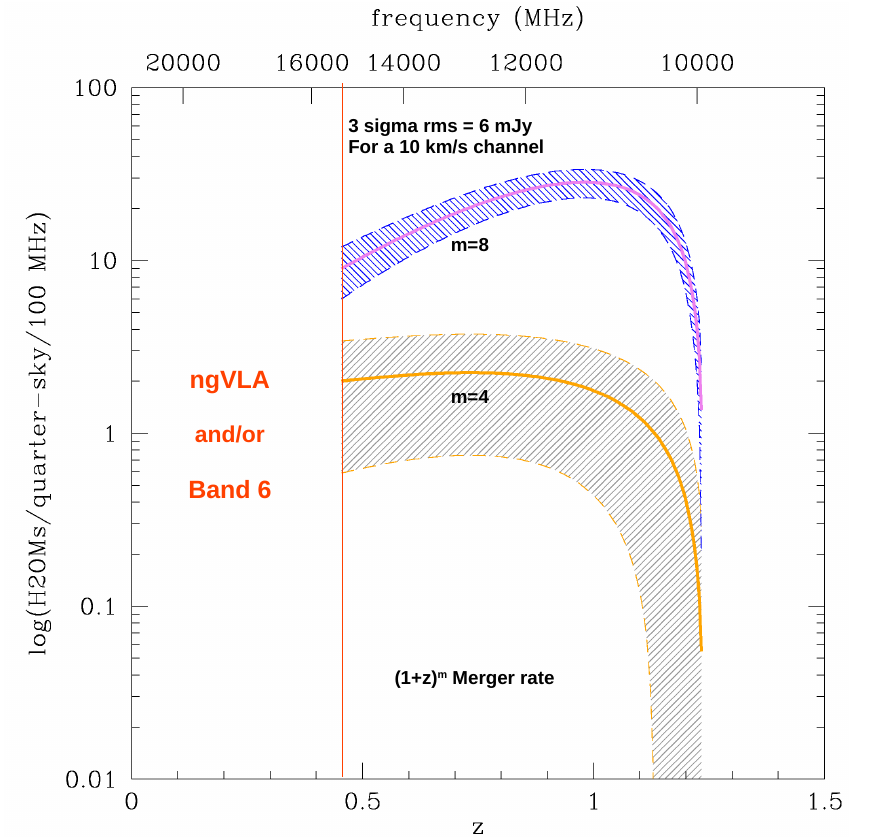}    \caption{Sky density of detectable water masers with the SKA-Mid telescope (AA4) blind-searching a quarter-sky area at a sensitivity of 6 mJy. Shaded areas indicate the 1$\sigma$ confidence level. The evolution of the LF with redshift is parametrized as $(1+z)^m$, where $m$ is 4 and 8 for moderate and very high evolution, respectively. The red solid line defines the actual lower limit in redshift ($z$ $\sim$ 0.45) for 22-GHz water maser studies possible with the SKA-Mid telescope due to the highest frequency threshold of band 5b (15.4 GHz).}
   \label{fig:halsky_figure}
\end{figure}

\begin{figure}[htbp]
    \centering
    \includegraphics[trim=120 420 120 120, clip, width=1.0\columnwidth]{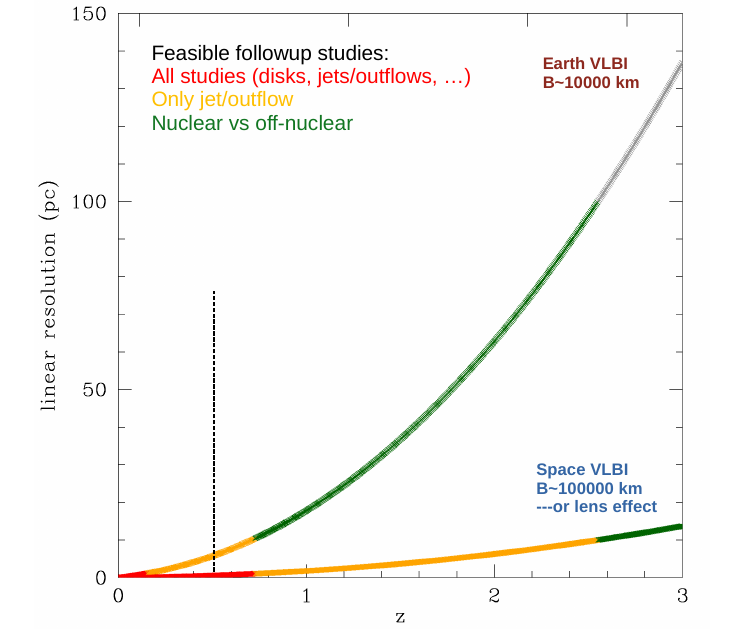}    
   \caption{Linear resolution as a function of redshift $z$ necessary to derive relevant information on the nuclear region of the host galaxies through water megamaser follow-up studies, for Earth-VLBI and Space-VLBI (or source lens-magnification) baselines, respectively. The color code is explained by the key appended in the top-left corner. The vertical black dashed-line marks the actual lower limit in redshift ($z$ $\sim$ 0.45) for 22-GHz water maser studies possible with the SKA-Mid telescope due to the highest frequency threshold of band 5b (15.4 GHz).}
   \label{fig:vlbi}
\end{figure}

\section{Acknowledgements}
We would like to thank Matteo Murgia for useful discussions and his support to improve the quality of the plots. 








\bibliographystyle{abbrvnat-maxbibnames4.bst}
\bibliography{Bibliografia_Tarchi_new} 

\end{document}

%% file: journal-names.tex
\newcommand{\actaa}{Acta Astron.} 
\newcommand{\araa}{ARA\&A} 
\newcommand{\aar}{A\&ARv} 
\newcommand{\aapr}{A\&ARv} 
\newcommand{\ab}{Astrobiol.} 
\newcommand{\aj}{AJ} 
\newcommand{\apj}{ApJ} 
\newcommand{\apjl}{ApJL} 
\newcommand{\apjs}{ApJSS} 
\newcommand{\ao}{Appl. Opt.} 
\newcommand{\apss}{Astro. \& Space Sci.} 
\newcommand{\aap}{A\&A} 
\newcommand{\aaps}{A\&AS.} 
\newcommand{\baas}{Bull. Am. Astron. Soc.} 
\newcommand{\caa}{Chinese A\&A} 
\newcommand{\cjaa}{Chinese J. A\&A} 
\newcommand{\cqg}{Class. Quantum Gravity} 
\newcommand{\gal}{Galaxies} 
\newcommand{\gca}{Geo. Cosmo. Acta} 
\newcommand{\icarus}{Icarus} 
\newcommand{\jcap}{JCAP} 
\newcommand{\jgr}{J. Geophys. Res.} 
\newcommand{\jgrp}{J. Geophys. Res. Planets} 
\newcommand{\jqsrt}{J. Quant. Spectrosc. Radiat. Transf.} 
\newcommand{\memsai}{Mem. SAIt} 
\newcommand{\mnras}{MNRAS} 
\newcommand{\nat}{Nature} 
\newcommand{\nastro}{Nat. Astron.} 
\newcommand{\ncomms}{Nat. Commun.} 
\newcommand{\nphys}{Nat. Phys.} 
\newcommand{\na}{New Astron.} 
\newcommand{\nar}{New Astron. Rev.} 
\newcommand{\physrep}{Phys. Rep.} 
\newcommand{\pra}{Phys. Rev. A} 
\newcommand{\prb}{Phys. Rev. B} 
\newcommand{\prc}{Phys. Rev. C} 
\newcommand{\prd}{Phys. Rev. D} 
\newcommand{\pre}{Phys. Rev. E} 
\newcommand{\prx}{Phys. Rev. X} 
\newcommand{\prl}{Phys. Rev. Let.} 
\newcommand{\psj}{Planet. Sci. J.} 
\newcommand{\planss}{Planet. Space Sci.} 
\newcommand{\pnas}{Proc. Natl Acad. Sci. USA} 
\newcommand{\procspie}{Proc. SPIE} 
\newcommand{\pasa}{PASA} 
\newcommand{\pasj}{PASJ} 
\newcommand{\pasp}{PASP} 
\newcommand{\rmxaa}{RMXAA} 
\newcommand{\sci}{Science} 
\newcommand{\sciadv}{Sci. Adv.} 
\newcommand{\solphys}{Sol. Phys.} 
\newcommand{\sovast}{Soviet Ast.} 
\newcommand{\ssr}{Space Sci. Rev.} 
\newcommand{\uni}{Universe} 

%% file: Bibliografia_Tarchi_new.bib
@ARTICLE{Greenhill2003,
       author = {{Greenhill}, L.~J. and {Booth}, R.~S. and {Ellingsen}, S.~P. and {Herrnstein}, J.~R. and {Jauncey}, D.~L. and {McCulloch}, P.~M. and {Moran}, J.~M. and {Norris}, R.~P. and {Reynolds}, J.~E. and {Tzioumis}, A.~K.},
        title = "{A Warped Accretion Disk and Wide-Angle Outflow in the Inner Parsec of the Circinus Galaxy}",
      journal = {\apj},
         year = 2003,
        month = jun,
       volume = {590},
       number = {1},
        pages = {162-173},
          doi = {10.1086/374862},
archivePrefix = {arXiv},
       eprint = {astro-ph/0302533},
 primaryClass = {astro-ph},
       adsurl = {https://ui.adsabs.harvard.edu/abs/2003ApJ...590..162G},
}

@INPROCEEDINGS{Herrnstein1997,
       author = {{Herrnstein}, J. and {Moran}, J. and {Greenhill}, L. and {Inoue}, M. and {Nakai}, N. and {Miyoshi}, M. and {Diamond}, P.},
        title = "{A 4\% Geometric Distance to NGC 4258 from Proper Motions in the Nuclear Water Maser}",
    booktitle = {American Astronomical Society Meeting Abstracts},
         year = 1997,
       series = {American Astronomical Society Meeting Abstracts},
       volume = {191},
        month = dec,
          eid = {25.07},
        pages = {25.07},
       adsurl = {https://ui.adsabs.harvard.edu/abs/1997AAS...191.2507H},
}

@ARTICLE{Kuo2020,
       author = {{Kuo}, C.~Y. and {Braatz}, J.~A. and {Impellizzeri}, C.~M.~V. and {Gao}, F. and {Pesce}, D. and {Reid}, M.~J. and {Condon}, J. and {Kamali}, F. and {Henkel}, C. and {Greene}, J.~E.},
        title = "{The Megamaser Cosmology Project - XII. VLBI imaging of H$_{2}$O maser emission in three active galaxies and the effect of AGN winds on disc dynamics}",
      journal = {\mnras},
         year = 2020,
        month = oct,
       volume = {498},
       number = {2},
        pages = {1609-1627},
          doi = {10.1093/mnras/staa2260},
       adsurl = {https://ui.adsabs.harvard.edu/abs/2020MNRAS.498.1609K},
}

@ARTICLE{Reid2009,
       author = {{Reid}, M.~J. and {Braatz}, J.~A. and {Condon}, J.~J. and {Greenhill}, L.~J. and {Henkel}, C. and {Lo}, K.~Y.},
        title = "{The Megamaser Cosmology Project. I. Very Long Baseline Interferometric Observations of UGC 3789}",
      journal = {\apj},
         year = 2009,
        month = apr,
       volume = {695},
       number = {1},
        pages = {287-291},
          doi = {10.1088/0004-637X/695/1/287},
archivePrefix = {arXiv},
       eprint = {0811.4345},
 primaryClass = {astro-ph},
       adsurl = {https://ui.adsabs.harvard.edu/abs/2009ApJ...695..287R},
}

@INPROCEEDINGS{Tarchi2012,
       author = {{Tarchi}, Andrea},
        title = "{AGN and Megamasers}",
    booktitle = {Cosmic Masers - from OH to H0},
         year = 2012,
       editor = {{Booth}, Roy S. and {Vlemmings}, Wouter H.~T. and {Humphreys}, Elizabeth M.~L.},
       volume = {287},
        month = jul,
        pages = {323-332},
          doi = {10.1017/S1743921312007259},
archivePrefix = {arXiv},
       eprint = {1205.3623},
 primaryClass = {astro-ph.CO},
       adsurl = {https://ui.adsabs.harvard.edu/abs/2012IAUS..287..323T},
}

@ARTICLE{Lo2005,
       author = {{Lo}, K.~Y.},
        title = "{Mega-Masers and Galaxies}",
      journal = {\araa},
         year = 2005,
        month = sep,
       volume = {43},
       number = {1},
        pages = {625-676},
          doi = {10.1146/annurev.astro.41.011802.094927},
       adsurl = {https://ui.adsabs.harvard.edu/abs/2005ARA&A..43..625L},
}

@ARTICLE{Panessa2020,
       author = {{Panessa}, F. and {Castangia}, P. and {Malizia}, A. and {Bassani}, L. and {Tarchi}, A. and {Bazzano}, A. and {Ubertini}, P.},
        title = "{Water megamaser emission in hard X-ray selected AGN}",
      journal = {\aap},
         year = 2020,
        month = sep,
       volume = {641},
          eid = {A162},
        pages = {A162},
          doi = {10.1051/0004-6361/201937407},
archivePrefix = {arXiv},
       eprint = {2006.08280},
 primaryClass = {astro-ph.GA},
       adsurl = {https://ui.adsabs.harvard.edu/abs/2020A&A...641A.162P},
}

@ARTICLE{Antonucci1993,
   author = {{Antonucci}, R.},
    title = "{Unified models for active galactic nuclei and quasars}",
  journal = {\araa},
     year = 1993,
   volume = 31,
    pages = {473-521},
      doi = {10.1146/annurev.aa.31.090193.002353},
   adsurl = {http://cdsads.u-strasbg.fr/abs/1993ARA\%26A..31..473A}
}

@ARTICLE{Bennert2009,
   author = {{Bennert}, N. and {Barvainis}, R. and {Henkel}, C. and {Antonucci}, R.
	},
    title = "{A Search for H$_{2}$O Megamasers in High-z Type-2 Active Galactic Nuclei}",
  journal = {\apj},
archivePrefix = "arXiv",
   eprint = {0901.0567},
 primaryClass = "astro-ph.GA",
     year = 2009,
    month = apr,
   volume = 695,
    pages = {276-286},
      doi = {10.1088/0004-637X/695/1/276},
   adsurl = {http://cdsads.u-strasbg.fr/abs/2009ApJ...695..276B}
}

@INPROCEEDINGS{Braatz2018,
   author = {{Braatz}, J. and {Condon}, J. and {Henkel}, C. and {Greene}, J. and 
	{Lo}, F. and {Reid}, M. and {Pesce}, D. and {Gao}, F. and {Impellizzeri}, V. and 
	{Kuo}, C.-Y. and {Zhao}, W. and {Constantin}, A. and {Hao}, L. and 
	{Litzinger}, E.},
    title = "{A Measurement of the Hubble Constant by the Megamaser Cosmology Project}",
booktitle = {Astrophysical Masers: Unlocking the Mysteries of the Universe},
     year = 2018,
   series = {IAU Symposium},
   volume = 336,
   editor = {{Tarchi}, A. and {Reid}, M.~J. and {Castangia}, P.},
    month = aug,
    pages = {86-91},
      doi = {10.1017/S1743921317010249},
   adsurl = {http://cdsads.u-strasbg.fr/abs/2018IAUS..336...86B}
}

@ARTICLE{Castangia2019,
       author = {{Castangia}, P. and {Surcis}, G. and {Tarchi}, A. and {Caccianiga}, A. and {Severgnini}, P. and {Della Ceca}, R.},
        title = "{Water masers in Compton-thick AGN. II. The high detection rate and EVN observations of <ASTROBJ>IRAS 15480-0344</ASTROBJ>}",
      journal = {\aap},
         year = 2019,
        month = sep,
       volume = {629},
          eid = {A25},
        pages = {A25},
          doi = {10.1051/0004-6361/201935421},
archivePrefix = {arXiv},
       eprint = {1907.09246},
 primaryClass = {astro-ph.GA},
       adsurl = {https://ui.adsabs.harvard.edu/abs/2019A&A...629A..25C}
}

@ARTICLE{Gao2017,
   author = {{Gao}, F. and {Braatz}, J.~A. and {Reid}, M.~J. and {Condon}, J.~J. and 
	{Greene}, J.~E. and {Henkel}, C. and {Impellizzeri}, C.~M.~V. and 
	{Lo}, K.~Y. and {Kuo}, C.~Y. and {Pesce}, D.~W. and {Wagner}, J. and 
	{Zhao}, W.},
    title = "{The Megamaser Cosmology Project. IX. Black Hole Masses for Three Maser Galaxies}",
  journal = {\apj},
archivePrefix = "arXiv",
   eprint = {1610.06802},
     year = 2017,
    month = jan,
   volume = 834,
      eid = {52},
    pages = {52},
      doi = {10.3847/1538-4357/834/1/52},
   adsurl = {http://cdsads.u-strasbg.fr/abs/2017ApJ...834...52G}
}

@ARTICLE{Henkel2005,
   author = {{Henkel}, C. and {Braatz}, J.~A. and {Tarchi}, A. and {Peck}, A.~B. and 
	{Nagar}, N.~M. and {Greenhill}, L.~J. and {Wang}, M. and {Hagiwara}, Y.
	},
    title = "{H$_{2}$O Megamasers: Accretion Disks, Jet Interaction, Outflows or Massive Star Formation?}",
  journal = {\apss},
   eprint = {astro-ph/0407161},
     year = 2005,
    month = jan,
   volume = 295,
    pages = {107-116},
      doi = {10.1007/s10509-005-3668-z},
   adsurl = {http://cdsads.u-strasbg.fr/abs/2005Ap%26SS.295..107H}
}

@ARTICLE{Kondratko2005,
   author = {{Kondratko}, P.~T. and {Greenhill}, L.~J. and {Moran}, J.~M.
	},
    title = "{Evidence for a Geometrically Thick Self-Gravitating Accretion Disk in NGC 3079}",
  journal = {\apj},
   eprint = {astro-ph/0408549},
     year = 2005,
    month = jan,
   volume = 618,
    pages = {618-634},
      doi = {10.1086/426101},
   adsurl = {http://cdsads.u-strasbg.fr/abs/2005ApJ...618..618K}
}

@ARTICLE{Kuo2011,
   author = {{Kuo}, C.~Y. and {Braatz}, J.~A. and {Condon}, J.~J. and {Impellizzeri}, C.~M.~V. and 
	{Lo}, K.~Y. and {Zaw}, I. and {Schenker}, M. and {Henkel}, C. and 
	{Reid}, M.~J. and {Greene}, J.~E.},
    title = "{The Megamaser Cosmology Project. III. Accurate Masses of Seven Supermassive Black Holes in Active Galaxies with Circumnuclear Megamaser Disks}",
  journal = {\apj},
archivePrefix = "arXiv",
   eprint = {1008.2146},
 primaryClass = "astro-ph.CO",
     year = 2011,
    month = jan,
   volume = 727,
      eid = {20},
    pages = {20},
      doi = {10.1088/0004-637X/727/1/20},
   adsurl = {http://adsabs.harvard.edu/abs/2011ApJ...727...20K}
}

@ARTICLE{Miyoshi1995,
   author = {{Miyoshi}, M. and {Moran}, J. and {Herrnstein}, J. and {Greenhill}, L. and 
	{Nakai}, N. and {Diamond}, P. and {Inoue}, M.},
    title = "{Evidence for a black hole from high rotation velocities in a sub-parsec region of NGC4258}",
  journal = {\nat},
     year = 1995,
    month = jan,
   volume = 373,
    pages = {127-129},
      doi = {10.1038/373127a0},
   adsurl = {http://cdsads.u-strasbg.fr/abs/1995Natur.373..127M}
}

@ARTICLE{Nenkova2008,
   author = {{Nenkova}, M. and {Sirocky}, M.~M. and {Nikutta}, R. and {Ivezi{\'c}}, {\v Z}. and 
	{Elitzur}, M.},
    title = "{AGN Dusty Tori. II. Observational Implications of Clumpiness}",
  journal = {\apj},
archivePrefix = "arXiv",
   eprint = {0806.0512},
     year = 2008,
    month = sep,
   volume = 685,
    pages = {160-180},
      doi = {10.1086/590483},
   adsurl = {http://cdsads.u-strasbg.fr/abs/2008ApJ...685..160N}
}

@ARTICLE{Peck2003,
   author = {{Peck}, A.~B. and {Henkel}, C. and {Ulvestad}, J.~S. and {Brunthaler}, A. and 
	{Falcke}, H. and {Elitzur}, M. and {Menten}, K.~M. and {Gallimore}, J.~F.
	},
    title = "{The Flaring H$_{2}$O Megamaser and Compact Radio Source in Markarian 348}",
  journal = {\apj},
   eprint = {astro-ph/0303423},
     year = 2003,
    month = jun,
   volume = 590,
    pages = {149-161},
      doi = {10.1086/374924},
   adsurl = {http://adsabs.harvard.edu/abs/2003ApJ...590..149P}
}

@ARTICLE{Almeida2017,
   author = {{Ramos Almeida}, C. and {Ricci}, C.},
    title = "{Nuclear obscuration in active galactic nuclei}",
  journal = {Nature Astronomy},
archivePrefix = "arXiv",
   eprint = {1709.00019},
     year = 2017,
    month = oct,
   volume = 1,
    pages = {679-689},
      doi = {10.1038/s41550-017-0232-z},
   adsurl = {http://adsabs.harvard.edu/abs/2017NatAs...1..679R}
}

@ARTICLE{Severgnini2012,
   author = {{Severgnini}, P. and {Caccianiga}, A. and {Della Ceca}, R.},
    title = "{A new technique to efficiently select Compton-thick AGN}",
  journal = {\aap},
archivePrefix = "arXiv",
   eprint = {1204.4359},
 primaryClass = "astro-ph.CO",
     year = 2012,
    month = jun,
   volume = 542,
      eid = {A46},
    pages = {A46},
      doi = {10.1051/0004-6361/201118417},
   adsurl = {http://cdsads.u-strasbg.fr/abs/2012A%26A...542A..46S}
}

@ARTICLE{Urry95,
   author = {{Urry}, C.~M. and {Padovani}, P.},
    title = "{Unified Schemes for Radio-Loud Active Galactic Nuclei}",
  journal = {\pasp},
   eprint = {astro-ph/9506063},
     year = 1995,
    month = sep,
   volume = 107,
    pages = {803},
      doi = {10.1086/133630},
   adsurl = {http://cdsads.u-strasbg.fr/abs/1995PASP..107..803U}
}

@ARTICLE{Braatz2010,
       author = {{Braatz}, J.~A. and {Reid}, M.~J. and {Humphreys}, E.~M.~L. and {Henkel}, C. and {Condon}, J.~J. and {Lo}, K.~Y.},
        title = "{The Megamaser Cosmology Project. II. The Angular-diameter Distance to UGC 3789}",
      journal = {\apj},
         year = 2010,
        month = aug,
       volume = {718},
       number = {2},
        pages = {657-665},
          doi = {10.1088/0004-637X/718/2/657},
archivePrefix = {arXiv},
       eprint = {1005.1955},
 primaryClass = {astro-ph.CO},
       adsurl = {https://ui.adsabs.harvard.edu/abs/2010ApJ...718..657B}
}

@ARTICLE{Bannikova2023,
       author = {{Bannikova}, Elena Yu and {Akerman}, Nina O. and {Capaccioli}, Massimo and {Berczik}, Peter P. and {Akhmetov}, Vladimir S. and {Ishchenko}, Marina V.},
        title = "{Apparent counter-rotation in the torus of NGC 1068: influence of an asymmetric wind}",
      journal = {\mnras},
         year = 2023,
        month = jan,
       volume = {518},
       number = {1},
        pages = {742-751},
          doi = {10.1093/mnras/stac3099},
archivePrefix = {arXiv},
       eprint = {2205.14455},
 primaryClass = {astro-ph.GA},
       adsurl = {https://ui.adsabs.harvard.edu/abs/2023MNRAS.518..742B}
}

@ARTICLE{Mckean2011,
       author = {{McKean}, J.~P. and {Impellizzeri}, C.~M.~V. and {Roy}, A.~L. and {Castangia}, P. and {Samuel}, F. and {Brunthaler}, A. and {Henkel}, C. and {Wucknitz}, O.},
        title = "{A search for gravitationally lensed water masers in dusty quasars and star-forming galaxies}",
      journal = {\mnras},
         year = 2011,
        month = feb,
       volume = {410},
       number = {4},
        pages = {2506-2515},
          doi = {10.1111/j.1365-2966.2010.17617.x},
archivePrefix = {arXiv},
       eprint = {1009.0290},
 primaryClass = {astro-ph.CO},
       adsurl = {https://ui.adsabs.harvard.edu/abs/2011MNRAS.410.2506M}
}

@ARTICLE{Greene2010,
       author = {{Greene}, Jenny E. and {Peng}, Chien Y. and {Kim}, Minjin and {Kuo}, Cheng-Yu and {Braatz}, James A. and {Impellizzeri}, C.~M.~V. and {Condon}, James J. and {Lo}, K.~Y. and {Henkel}, Christian and {Reid}, Mark J.},
        title = "{Precise Black Hole Masses from Megamaser Disks: Black Hole-Bulge Relations at Low Mass}",
      journal = {\apj},
         year = 2010,
        month = sep,
       volume = {721},
       number = {1},
        pages = {26-45},
          doi = {10.1088/0004-637X/721/1/26},
archivePrefix = {arXiv},
       eprint = {1007.2851},
 primaryClass = {astro-ph.CO},
       adsurl = {https://ui.adsabs.harvard.edu/abs/2010ApJ...721...26G}
}

@ARTICLE{Barvainis2005,
       author = {{Barvainis}, Richard and {Antonucci}, Robert},
        title = "{Extremely Luminous Water Vapor Emission from a Type 2 Quasar at Redshift z = 0.66}",
      journal = {\apjl},
         year = 2005,
        month = aug,
       volume = {628},
       number = {2},
        pages = {L89-L91},
          doi = {10.1086/432666},
archivePrefix = {arXiv},
       eprint = {astro-ph/0506245},
 primaryClass = {astro-ph},
       adsurl = {https://ui.adsabs.harvard.edu/abs/2005ApJ...628L..89B}
}

@ARTICLE{Impellizzeri2008,
       author = {{Impellizzeri}, C.~M. Violette and {McKean}, John P. and {Castangia}, Paola and {Roy}, Alan L. and {Henkel}, Christian and {Brunthaler}, Andreas and {Wucknitz}, Olaf},
        title = "{A gravitationally lensed water maser in the early Universe}",
      journal = {\nat},
         year = 2008,
        month = dec,
       volume = {456},
       number = {7224},
        pages = {927-929},
          doi = {10.1038/nature07544},
archivePrefix = {arXiv},
       eprint = {0901.1132},
 primaryClass = {astro-ph.CO},
       adsurl = {https://ui.adsabs.harvard.edu/abs/2008Natur.456..927I}
}

@ARTICLE{Omont2013,
       author = {{Omont}, A. and {Yang}, C. and {Cox}, P. and {Neri}, R. and {Beelen}, A. and {Bussmann}, R.~S. and {Gavazzi}, R. and {van der Werf}, P. and {Riechers}, D. and {Downes}, D. and {Krips}, M. and {Dye}, S. and {Ivison}, R. and {Vieira}, J.~D. and {Wei{\ss}}, A. and {Aguirre}, J.~E. and {Baes}, M. and {Baker}, A.~J. and {Bertoldi}, F. and {Cooray}, A. and {Dannerbauer}, H. and {De Zotti}, G. and {Eales}, S.~A. and {Fu}, H. and {Gao}, Y. and {Gu{\'e}lin}, M. and {Harris}, A.~I. and {Jarvis}, M. and {Lehnert}, M. and {Leeuw}, L. and {Lupu}, R. and {Menten}, K. and {Micha{\l}owski}, M.~J. and {Negrello}, M. and {Serjeant}, S. and {Temi}, P. and {Auld}, R. and {Dariush}, A. and {Dunne}, L. and {Fritz}, J. and {Hopwood}, R. and {Hoyos}, C. and {Ibar}, E. and {Maddox}, S. and {Smith}, M.~W.~L. and {Valiante}, E. and {Bock}, J. and {Bradford}, C.~M. and {Glenn}, J. and {Scott}, K.~S.},
        title = "{H$_{2}$O emission in high-z ultra-luminous infrared galaxies}",
      journal = {\aap},
         year = 2013,
        month = mar,
       volume = {551},
          eid = {A115},
        pages = {A115},
          doi = {10.1051/0004-6361/201220811},
archivePrefix = {arXiv},
       eprint = {1301.6618},
 primaryClass = {astro-ph.CO},
       adsurl = {https://ui.adsabs.harvard.edu/abs/2013A&A...551A.115O}
}

@ARTICLE{Darling2002,
       author = {{Darling}, Jeremy and {Giovanelli}, Riccardo},
        title = "{The OH Megamaser Luminosity Function}",
      journal = {\apj},
         year = 2002,
        month = jun,
       volume = {572},
       number = {2},
        pages = {810-822},
          doi = {10.1086/340372},
archivePrefix = {arXiv},
       eprint = {astro-ph/0204195},
 primaryClass = {astro-ph},
       adsurl = {https://ui.adsabs.harvard.edu/abs/2002ApJ...572..810D}
}

@ARTICLE{Tarchi2007,
       author = {{Tarchi}, A. and {Brunthaler}, A. and {Henkel}, C. and {Menten}, K.~M. and {Braatz}, J. and {Wei{\ss}}, A.},
        title = "{The innermost region of the water megamaser radio galaxy 3C 403}",
      journal = {\aap},
         year = 2007,
        month = nov,
       volume = {475},
       number = {2},
        pages = {497-506},
          doi = {10.1051/0004-6361:20078317},
archivePrefix = {arXiv},
       eprint = {0709.3417},
 primaryClass = {astro-ph},
       adsurl = {https://ui.adsabs.harvard.edu/abs/2007A&A...475..497T}
}

@ARTICLE{Castangia2011,
       author = {{Castangia}, P. and {Impellizzeri}, C.~M.~V. and {McKean}, J.~P. and {Henkel}, C. and {Brunthaler}, A. and {Roy}, A.~L. and {Wucknitz}, O. and {Ott}, J. and {Momjian}, E.},
        title = "{Water vapour at high redshift: Arecibo monitoring of the megamaser in MG J0414+0534}",
      journal = {\aap},
         year = 2011,
        month = may,
       volume = {529},
          eid = {A150},
        pages = {A150},
          doi = {10.1051/0004-6361/201016403},
archivePrefix = {arXiv},
       eprint = {1103.4301},
 primaryClass = {astro-ph.CO},
       adsurl = {https://ui.adsabs.harvard.edu/abs/2011A&A...529A.150C}
}

@ARTICLE{Surcis2009,
       author = {{Surcis}, G. and {Tarchi}, A. and {Henkel}, C. and {Ott}, J. and {Lovell}, J. and {Castangia}, P.},
        title = "{New H2O masers in Seyfert and FIR bright galaxies. III. The southern sample}",
      journal = {\aap},
         year = 2009,
        month = aug,
       volume = {502},
       number = {2},
        pages = {529-540},
          doi = {10.1051/0004-6361/200911685},
archivePrefix = {arXiv},
       eprint = {0904.4816},
 primaryClass = {astro-ph.CO},
       adsurl = {https://ui.adsabs.harvard.edu/abs/2009A&A...502..529S}
}

@INPROCEEDINGS{Pesce2018,
       author = {{Pesce}, Dominic and {Braatz}, James and {Condon}, James and {Gao}, Feng and {Henkel}, Christian and {Impellizzeri}, Violette and {Litzinger}, Eugenia and {Lo}, K.~Y. and {Reid}, Mark},
        title = "{AGN accretion disk physics using H$_{2}$O megamasers}",
    booktitle = {Astrophysical Masers: Unlocking the Mysteries of the Universe},
         year = 2018,
       editor = {{Tarchi}, A. and {Reid}, M.~J. and {Castangia}, P.},
       series = {IAU Symposium},
       volume = {336},
        month = aug,
        pages = {125-128},
          doi = {10.1017/S1743921317009966},
       adsurl = {https://ui.adsabs.harvard.edu/abs/2018IAUS..336..125P}
}

@ARTICLE{Tarchi2024,
       author = {{Tarchi}, A. and {Castangia}, P. and {Surcis}, G. and {Impellizzeri}, V. and {Ladu}, E. and {Bannikova}, E. Yu.},
        title = "{Detection of maser emission at 183 and 380 GHz with ALMA in the gigamaser galaxy TXS 2226-184}",
      journal = {\aap},
         year = 2024,
        month = aug,
       volume = {688},
          eid = {L18},
        pages = {L18},
          doi = {10.1051/0004-6361/202451245},
archivePrefix = {arXiv},
       eprint = {2407.13373},
 primaryClass = {astro-ph.GA},
       adsurl = {https://ui.adsabs.harvard.edu/abs/2024A&A...688L..18T}
}

@incollection{Rygl01.2026.SKA, author = {Kazi L. J. Rygl and author2 and author3 and author4 and author5},title = {},year = {2026},publisher = {},note = {arXiv search: Report number AASKAII/Rygl01},booktitle = {Advancing Astrophysics with the SKA -- II (AASKAII)}}
